# NETWORK CALCULUS BASED FDI APPROACH FOR SWITCHED ETHERNET ARCHITECTURE

B. Brahimi, C. Aubrun, E. Rondeau

CRAN/UHP/CNRS UMR 7039
Faculté des Sciences et Techniques
BP 239
54506 Vandoeuvre-lès-Nancy Cedex
France

Abstract : The Networked Control Systems (NCS) are complex systems which integrate information provided by several domians such as automatic control, computer science, communication network. The work presented in this paper concerns fault detection, isolation and compensation of communication network. The proposed method is based on the classical approach of Fault Detection and Isolation and Fault Tolerant Control (FDI/FTC) currently used in diagnosis. The modelling of the network to be supervised is based on both couloured petri nets and network calculus theory often used to represent and analyse the network behaviour. The goal is to implement inside network devices algorithms enabling to detect, isolate and compensate communication faults in an autonomous way.

## 1. INTRODUCTION

The Network Control Systems (NCS) consist in controllers, sensors, actuators, which communicate thru the network (Zhang, *et al*, 2001; Tipsuwan and Chow, 2003). NCS are used in many industrial applications as they reduce wiring costs, improve the performances of the system diagnosis and the maintenance, and allows modularity and flexibility in the system design.

The implementation of network inside the control loop of the system makes complex the NCS analysis. Conventional control theory under restrictive assumptions, such as synchronized control and non delayed sensing and actuation, must be re-evaluated before being applied to the NCS. The main problems induced by the network are delay, jitter due to the delay variation, packet loss, (Zhang, *et al*., 2001). These troubles may produce instability of the distributed control systems (Juanole, *et al*., 2005).

The goal of the design of NCS is to improve the network performances in order to minimize the communication impact on the remote control. Thus, it is important to have a good knowledge of the communication system to achieved in real-time diagnosis. Network faults which generate missing values, increase delays, etc…should be detected and isolated. Then, mechanisms which enable fault effect compensation by changing the network routes, the message priorities, dropping non real-time packets,

etc…, should be triggered. Currently, the supervision of the network is centralized. It is necessary to send SNMP packets which tend to overload or to produce additional traffic to capture the state of the communication system. The objective of the work presented in this paper is to suppress partially the network management traffic by implementing directly in the network devices FDI and FTC algorithms.

The networks currently used in industrial systems are the fieldbuses, such as FIP, Profibus, DeviceNet, ControlNet. These networks have been specified to respect the temporal constraints defined in the industrial applications. However, their high price, their limited bandwidth and the problem of interoperability between these networks and the backbone network of the enterprise are recurrent difficulties affecting the integration of the fieldbuses (Ji and Kim 2005).

Several researches (Lee and Lee, 2002; Ji and Kim, 2005, Georges, *et al.*, 2005) show the efficiency of Ethernet network for industrial environment. The advantages of Ethernet in industry lies in a open and flexible communication system, independent from specific technologies (PLC, Supervisor,…) and well known by network engineers.

The objective of this paper is to propose FDI/FTC algorithms to be implemented in Ethernet switches. These algorithms are based on network calculus theory and Coloured Petri Net (CPN) models.

## 2. NETWORK FAULTS

*2.1 FAULT DEFINITION*

The fault defined in (Arlat, *et al.*, 2005; Avizienis, et al., 2004) is associated to a cause which corresponds to an error. Correct service is delivered when the service implements the system function. A service failure, often abbreviated failure, is an event that occurs when the delivered service deviates from correct service. The consequence of failure is a fault.

Since a service is a sequence of the system's external states, a service failure means that at least one (or more) external state of the system deviates from the correct service state. The deviation is called an error. Few works take into account the network faults; several authors such as (Feather, *et al.*, 1993; Pencolé, 2002) define tow kind of failures in the network: hardware failures and software failures.

The hard failures are characterized by the inability to deliver packets, while the soft failures are characterized by a partial loss, or delay of packets or a loss of a network bandwidth. In the two cases the symptom of these failures is the delay (this delay can be constant or variable, important or insignificant). Possible causes of a hard failure include cut cable, or failure of major network equipment (e.g. switch failure).

Some related works (Lelevé, 2000; Laprie, 2004; Pencolé, 2002; Aubrun, *et al.*, 2005) agrees with the fact that possible causes of soft failures are characterized by degraded performance such as delay, packet loss or network bandwidth reduction. Other type of failures may be caused by inappropriate network configuration (changes in network parameters) and also to new traffic generated by unknown applications. All these problems may provoke congestion and then may increase delay in a significant way for the real-time applications. In this paper we will focus, mainly, on the soft failure, and more specifically the objective is to detect an anomaly on the traffic variation.

*2.2 FAULT DETECTION AND ISOLATION*

In this work an Isolation and Detection of fault are carried out by detecting symptom (the error) (Laprie, 2001). The first step of the detection consists in estimating for each, real-time traffic their maximum delay to cross a switch in a normal functioning.

It is considered that these bounded delays have to be respected to guarantee that all industrial devices (PLC,…) receive remote information in a temporal window predefined by the application constraints. These constraints are called the Quality of Service (QoS) required by the control system with the capability to adapt this QoS by the QoS provided by the network. Thus, in the Ethernet switch, two procedures are implementing. The first one estimates the maximum delay to cross this switch for a given flow. This estimation is achieved by using the network calculus theory (Georges, 2005; Georges, *et al.*, 2005). The second one measures on line these delays. When the measure is superior to the estimated value, a fault is automatically detected.

*2.3 FAULT COMPENSATION*

In a switch, different mechanisms of compensation can be applied. These mechanisms depend on the technology of the switch. But the main actions for compensating a fault inside a switch are to change the priority tag on the frames, to modify the scheduling rules on the queues of the switches, to change the bandwidth, to drop packets,…These compensation mechanisms are invoked according to the importance of the fault on the traffic. The graduation of the fault effect depends on the temporal constraint associated to the frame identified for example by the priority tag. It depends also on the level of deviation observed during the fault detection step.

## 3. NETWORK MODELLING

### 3.1 ETHERNET SWITCH MODELLING

A Switch is a complex system which includes different mechanisms and technologies. (Philipps, 1999) and (Seifert, 2000) decompose the switching architecture in three main functional components:
- The queuing models refers to the buffering and the congestion mechanisms located in the switch,
- The switching implementation refers to the decision making process within the switch (how and where a switching decision is made),
- And the switching fabric is the path that data take to move from one port to another. There are different ways to build each of these components into a switching architecture. In this paper, only one typical kind of switching architecture is considered: Cisco Catalyst 2900 XL. Its model is shown on the figure 1. It is constituted of a sequence of multiple components:
- One multiplexer and one queue to represent a switch using a shared memory,
- One demultiplexer to model the switching step,
- One demultiplexer for each output port
- As much buffer as defined priority (until 8) for each output port,
- One multiplexer for each output port defining the bandwidth used.
The figure 1 shows the model of a switch which manages two priorities on the frames.

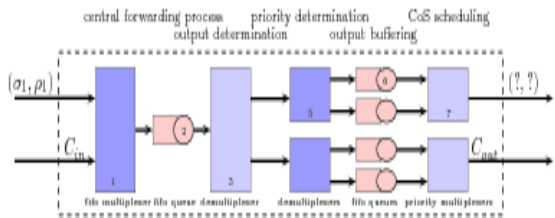

Fig. 1. IEEE 802.1p/Q Switch model.

The objective of this work is to propose a general method which integrates FDI/FTC procedures inside an Ethernet switch. This system is modelled by using CPN (Jensen, 1992)

The procedure presented in this paper is divided into three steps :
- In first step, CPN is used to model the Ethernet switch shown in the figure 1. The goal is to obtain simulated delay with CPN which corresponds to the measured value of the delay in the FDI procedure,
- In a second step, CPN is used to model the fault detection by comparing the estimated value obtained by the network calculus theory and the measured value obtained from CPN simulation.
- In the third step, CPN is used to model fault compensation on the traffic. The modelling was carried out with a modelling, simulation and verification tool: CPN Tools

### 3.2 ETHERNET SWITCH MODELLING WITH CPN

The switch depicted in figure 1 is modelled by using the hierarchical temporal coloured Petri Nets. The switch comprises three levels: 1, 2 and 3 (figure 2 in appendix ). At first, it is represented by a substitute transition (Kristensen, *et al.*, 1998), which represents the higher level of the hierarchy (level 3). This transition is divided in four modules (level 2 of the figure 2). The first module represents the FIFO queue, the second one represents the set of demultiplexers, and the third one represents the FIFO queues set associated to each output port of demultiplexers. Finally, and the last module represents scheduling priority mechanism. At the level 1 each module is modelled by coloured Petri nets, the communication between a module with another one is achieved via common places. The mechanism of priority scheduling selected for this modelling is a static priority. In a first approach, as it is the main traffic in NCS, only periodic communications are modelled.

## 4. FDI MODELLING

Figure 3 represents the FDI module. In this configuration three level of priority of packets are considered (3: high priority (HH), 2: mean priority (HB), 1: low priority (BH)). The places PH1, PM1 and PB1 represent the normal state of switch. Indeed, the marking of these places means that the delay of the different packets is lower to the theoretical end to end delay calculated by the network calculus. The places PH, PM and PB represent abnormal state of switch, the marking of one of them means that the delay of one of the different packet is superior to the theoretical end to end delay

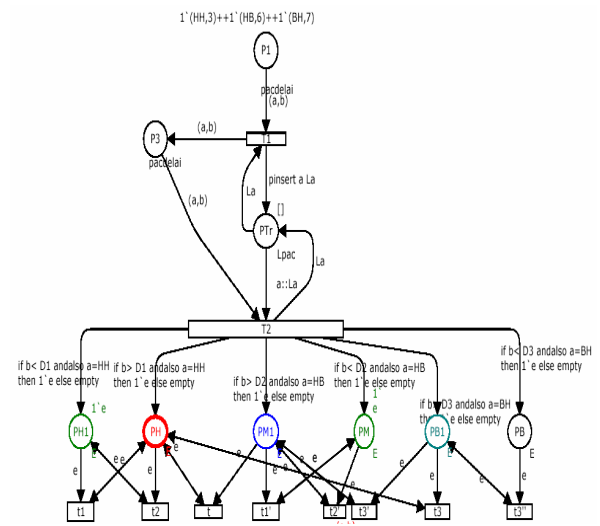

Fig. 3. FDI module

When the delay is detected as a fault, a compensation procedure is executed.

## 5. FTC MODELLING

For the compensation procedure, a method enabling the improvement of the quality of service is proposed. The general objective of the FTC procedure is to route the faulty packets in the queues with high priority. This routing takes into account the priority associated to each packet when faults occur in packets with different priorities at same time. The figure 4 shows the FTC procedure based on CPN modelling.

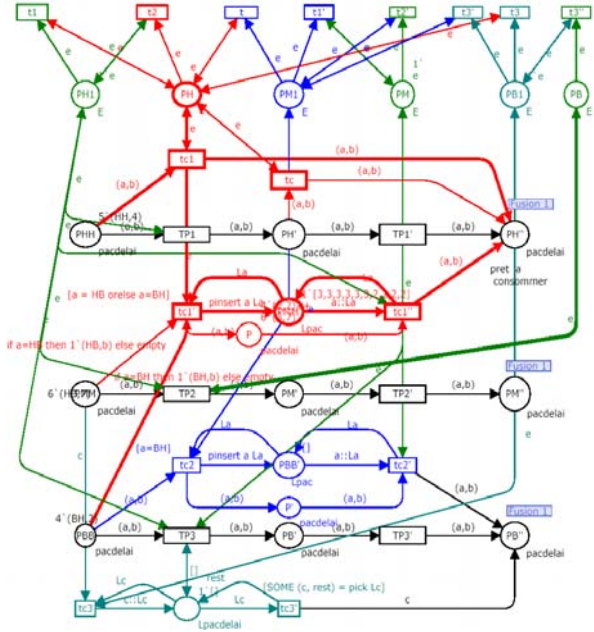

Fig. 4. FTC module

*FTC Algorithm*
The proposed FTC Algorithm is described as follows:
*if* $D_{HP}$ (delay of the high priority packet) > $DBB_T$ (End to end theoretical delay) *then*

    *if* $D_{MP}$ (delay of the mean priority packet ) > $DBB_T$ *then*

        *if* $D_{BP}$ (delay of the low priority packet) > $DBB_T$ *then*
        1-set $Packet_{MP}$ et $packet_{BP}$ in FIFO queue (priority-FIFO) ;
        2-transmit the $Packet_{HP}$ ;
    *Else*
        Action 1 and 2
    *Endif*
*Else*
    Action 1 and 2
*Endif*

*Else if* ($D_{MP}$ < $DBB_T$ and $D_{BP}$ > $DBB_T$) *then*
    Transmit $Packet_{BM}$ if there is not $Packet_{MP}$ to trasmit (Zero test)
    *Else*
        1-Set $Packet_{BM}$ in FIFO queue
        2-Transmit the $Packet_{HP}$
        3-Transmit the $Packet_{MP}$
    *Endif*
*Endif*

This algorithm is modelled by the CPN presented at the Figure 4. PHH, PMM and PBB represent the places which contain respectively, the high, mean, and low priority Packets, these places are in the output of FIFO queue which are in the output of each demultiplexer port.

## 6. SIMULATION AND RESULTS

### 6.1 MODEL SIMULATION

for the simulation of the model presented above, the following parameters are set:
- a period of traffic source is 5 T.U[1],
- consumer model 2 T.U.

The traffic source (which is, in fact, four sources packed on one), generate four kind of packets: packets crossing the switch from the input port 1(2) to the output port1 (2) - for example a source generates a packet which goes from the input port 1 to the output port1 with a high priority. One second source generates a packet which crosses the switch from the input port 1 to the output port 1 with a low priority, and the same behaviour for the remaining sources. The model is simulated with 10 000 steps of simulation. One can notice that the end to end delays of the traffic varies between 4 T.U and 2 T.U (see figure 5-a/). Figure 5-b/ represents the case where about 3 packets per period are added in order to load the switch. The purpose of this additional traffic is to show the consequences of a software fault on the switch behaviour. The fault considered here can be due to downloaded data or a sensor failure, which has as a consequence traffic congestion within the switch. It should be noted, that here we assume that there are no physical faults, and that the band-width is constant.

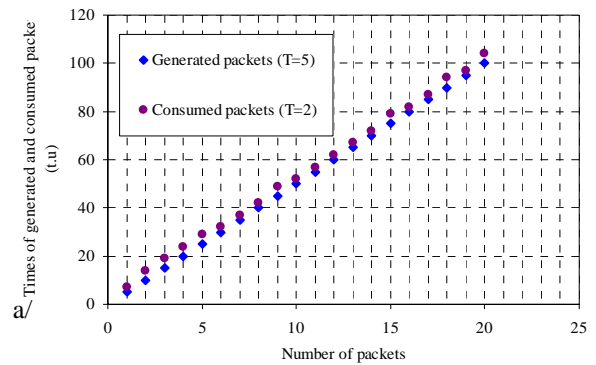

a/

---

[1] T.U: Time Unit, means all the parameters have the same time unit, for example: if T.U = ms then all parameters are ms.

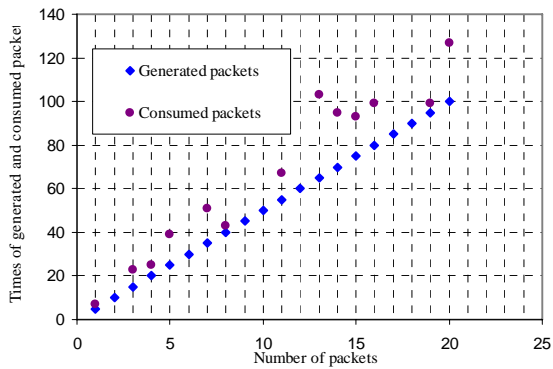

Fig. 5. End to end delay of high priority packets: a/ of switch without congestion; b/ to a switch with congestion

## 6.2 RESULTS OF SIMULATIONS

The parameters of simulation are identical to the previous simulation. However, here there are 6 traffic sources packed bended on one. Three different priorities are associated to different packets (HP, MP, BP). A burst of 3 packets per period leads to switch congestion which causes unexpected delay. The FDI module detects this delay on the base of the comparison between the simulation delay and theoretical end to end delay $DBB_T = 80$ T.U. The proposed FTC algorithm described by the scheme at figure 4 is applied to the Ethenet switch figure 6/b represents the behaviour of the end to end delays. One can notice that the $10^{th}$ packet; is not represented as its delay exceeds the value of $DBB_T$. The FDI module detects the delays occurring for various packets. In this case, the switch changes from a configuration towards an adapted scheduling policy mechanism. The aim of this mechanism is to transmit the higher priority packets without delay. The results depicted at figure 6.b/, show that the high priority packets are consumed regularly, according to the theoretical end to end delay, which proves a correct functioning of the switch.

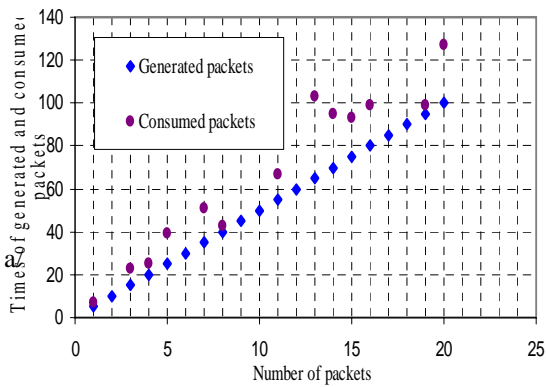

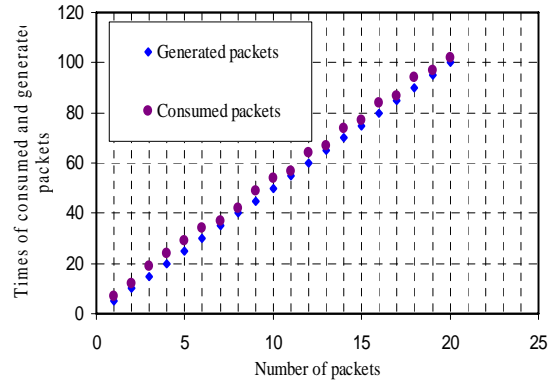

Fig. 6. Results for 10 000 simulation step, a/ high priority packets with congestion,. b/ high priority packets after compensation.

## 7. CONCLUSION

In this paper fault detection and accommodation procedure for switch Ethernet has been presented. First, a model of the switch has been presented, and then a fault detection and accommodation has been proposed. This work deals with communication network aspect: At first, the soft fault detection method is presented, it is carried out with use of end to end delay calculated by network calculus. This value is used at threshold witch is compared to the simulated value of the delay. The model of the switch is based on CPN modelling. The goal here is to show that the combination between the quantitative and qualitative modelling provides promising result for FDI and FTC for network components.

## ACKNOWLEDGMENT

This research has been conducted as a part of the Networked Control Systems Tolerant to faults (NeCST) Project contract n° IST – 2004-004303.

APPENDIX

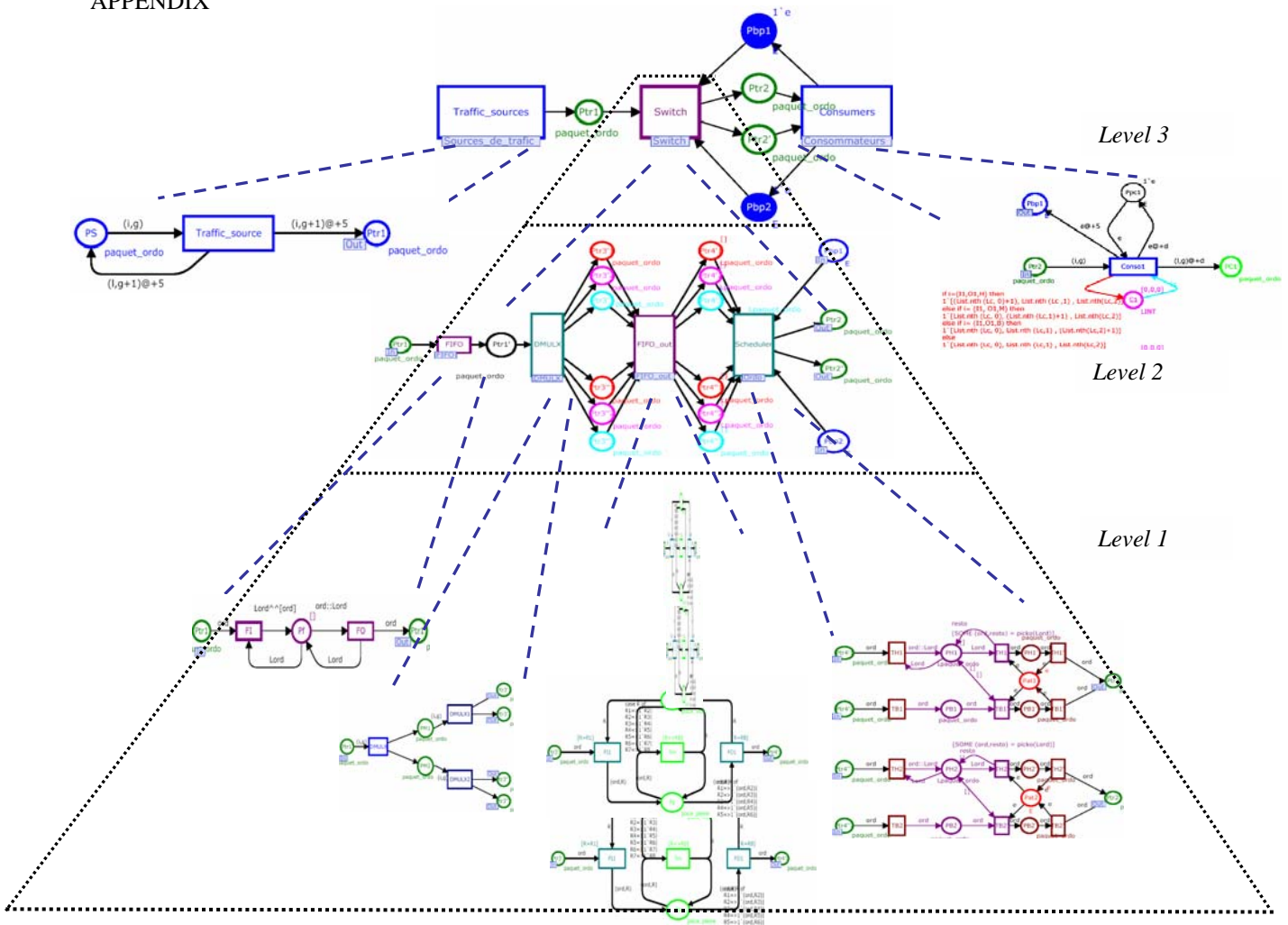

Fig. 2. Hierarchical model of Ethernet switch